\documentclass[11pt]{article}
\usepackage{amssymb,amsmath,fullpage}
\usepackage{color}
\numberwithin{equation}{section}
\newtheorem{theorem}{Theorem}[section]
\newtheorem{proposition}[theorem]{Proposition}
\newtheorem{lemma}[theorem]{Lemma}
\newtheorem{remark}[theorem]{Remark}

\newcommand{\proof}[1]{\par\smallskip\noindent{\em Proof #1}:\ }
\newcommand{\qed}{\ \hfill $\square$\par\medskip}
\newcommand{\dsum}[2]{{\displaystyle\sum_{#1}^{#2}}}

\newcommand{\dprod}[2]{{\displaystyle\prod_{#1}^{#2}}}
\newcommand{\tprod}[2]{{\textstyle\prod_{#1}^{#2}}}
\newcommand{\pr}[1]{\left\{{#1}\right\}}
\newcommand{\hf}{\frac{1}{2}}
\renewcommand{\Im}{\mbox{\rm Im}}

\title{\bf An infinite family of higher-order difference operators that 
commute with Ruijsenaars operators of type $A$} 
\author{Masatoshi NOUMI\footnote{
Department of Mathematics, KTH Royal Institute of Technology; 
SE-100 44 Stockholm, Sweden \hfill\break
(on leave from: Department of Mathematics, Kobe University; 
Rokko, Kobe 657-8501, Japan)}
\ \ 
and Ayako SANO\footnote{
Department of Mathematics, Kobe University (graduate); 
Rokko, Kobe 657-8501, Japan
}
}

\date{}
\begin{document}
\maketitle

%%%%%%%%%%%%%%%%%%%%%%%%%%%%%%%%%%%%%%

\begin{quote}
{\bf Abstract:}\quad We introduce a new infinite family of higher-order difference 
operators that commute with the elliptic Ruijsenaars difference 
operators of type $A$. 
These operators are related with Ruijsenaars' operators through 
a formula of Wronski type.
% which is proved by a functional identity for the sigma function. 
\par\medskip
{\em Keywords\/}: Ruijsenaars operators, Wronski formula
\newline
{\em 2010 Mathematical Subject Classification\/}:\  81R12; 33D67
\end{quote}

%%%%%%%%%%%%%%%%%%%%%%%%%%%%%%%%%%%%%%

\section{Introduction}

In his study of relativistic quantum integrable systems \cite{Ru1987}, 
Ruijsenaars introduced a commuting family of linear 
difference operators in $n$ variables, denoted below by $D_1,\ldots, D_n$,  
involving sigma functions as coefficients. 
In this paper 
we construct an explicit infinite family of difference operators 
$H_0, H_1, H_2,\ldots$ 
in the commutative algebra $\mathbb{C}[D_1,\ldots,D_n]$ 
which are related with $D_r$ 
($r=1,\ldots,n$) through a formula of Wronski type. 
This construction is applicable also to difference operators with 
trigonometric and rational coefficients. 
In order to deal with the elliptic, trigonometric and rational cases simultaneously, 
as in \cite{KN2003} and \cite{KNS2009}
we formulate our results in terms of an entire function $[z]$ satisfying 
the three term relation \eqref{eq:Hirota} below. 

\par\medskip
We fix a nonzero entire function $[z]$ in one complex variable $z\in\mathbb{C}$ 
satisfying the three term relation of Hirota type
\begin{equation}\label{eq:Hirota}
[z\pm \alpha][\beta\pm\gamma]+[z\pm\beta][\gamma\pm\alpha]+[z\pm\gamma][\alpha\pm\beta]=0
\end{equation}
for any $\alpha,\beta,\gamma\in\mathbb{C}$, where 
$[\alpha\pm\beta]=[\alpha+\beta][\alpha-\beta]$. 
We remark 
that a generic solution of this functional equation is given by
\begin{equation}
[z]=\mbox{const.}\,e^{cz^2}\sigma(z;\Omega)\qquad(c\in\mathbb{C}), 
\end{equation}
where $\sigma(z;\Omega)$ denotes the Weierstrass sigma function 
associated with a period lattice $\Omega=\mathbb{Z}\omega_1\oplus\mathbb{Z}\omega_2$. 
It is also satisfied by the functions
\begin{equation}
[z]=\mbox{const.}\,e^{cz^2}\sin(\pi z/\omega),\quad 
[z]=\mbox{const.}\,e^{cz^2}\,z,
\end{equation}
which are trigonometric and rational degenerations of the generic solution above. 
It is known that any solution of \eqref{eq:Hirota} belongs to one of these 
three categories. 
Throughout this paper, 
we denote by $D_r=D_r^{x}$ ($r=1,\ldots,n$) 
the Ruijsenaars operators in $n$ variables $x=(x_1,\ldots,x_n)$ 
with parameters $(\delta,\kappa)$, associated with $[z]$. 
They are defined by 
\begin{equation}
D_{r}^{x}=\sum_{I\subseteq \{1,\ldots,n\};\,|I|=r}\ 
\dprod{i\in I;\ j\notin I}{}\dfrac{[x_i-x_j+\kappa]}{[x_i-x_j]}\,
\dprod{i\in I}{}\,T_{x_i}^{\delta},
\end{equation}
where $I$ runs over all subsets of indices of cardinality $r$, 
and $T_{x_i}^{\delta}$ stands for the $\delta$-shift operator 
$x_i\to x_i+\delta$ in $x_i$ for each $i=1,\ldots,n$. 
It is proved by Ruijsenaars \cite{Ru1987} that these operators $D_{r}$ 
commute with each other, 
namely, 
\begin{equation}
D_rD_s=D_sD_r\qquad\mbox{for all $r,s=1,\ldots,n$,}
\end{equation}
on the basis of a certain functional identity for the sigma function. 
We denote by 
$\mathcal{R}=\mathbb{C}[D_1,\ldots,D_n]$ 
the commutative algebra generated by $D_{r}$ ($r=1,\ldots,n$), 
and refer to it as the commutative algebra of 
{\em Ruijsenaars operators} (of type $A_{n-1}$).  
We also define $D_{0}=1$, and $D_{r}=0$ for $r>n$. 
Note also that these operators 
$D_{r}$ ($r=1,\ldots,n$) in the trigonometric case 
are the Macdonald $q$-difference operators expressed additively. 

\par\medskip
We define an infinite family of difference operators 
$H_l=H_l^{x}$ $(l=0,1,2\ldots)$ by
\begin{equation}\label{eq:defH}
H_l^{x}
=\dsum{\mu_1+\cdots+\mu_n=l}{}\ 
\dprod{1\le i<j\le n}{}\!\dfrac{[x_i\!-\!x_j\!+\!(\mu_i\!-\!\mu_j)\delta]}
{[x_i\!-\!x_j]}\,
\dprod{i=1}{n}\dprod{j=1}{n}\dprod{k=0}{\mu_i\!-\!1}
\dfrac{[x_i\!-\!x_j\!+\!\kappa\!+\!k\delta]}{[x_i\!-\!x_j\!+\!\delta\!+\!k\delta]}\,
\dprod{i=1}{n}\,T_{x_i}^{\mu_i\delta}. 
\end{equation}
These operators are expressed briefly as 
\begin{equation}
H_l^{x}=
\dsum{\mu\in\mathbb{N}^n;\,|\mu|=l}{}
\dfrac{\Delta(x+\mu\delta)}{\Delta(x)}\,
\dprod{i,j=1}{n}\dfrac{[x_i-x_j+\kappa]_{\mu_i}}{[x_i-x_j+\delta]_{\mu_i}}\,T_{x}^{\mu\delta},
\end{equation}
in terms of the difference product 
%\begin{equation}
$\Delta(x)=\dprod{1\le i<j\le n}{}[x_i-x_j]$,  
%\end{equation}
and the $\delta$-shifted factorials 
\begin{equation}
[z]_k=[z][z+\delta]\cdots[z+(k-1)\delta]\quad(k=0,1,2,\ldots) 
\end{equation}
for $[z]$. 
Also, for each multi-index 
$\mu=(\mu_1,\ldots,\mu_n)\in\mathbb{N}^n$, $\mathbb{N}=\{0,1,2,\ldots\}$, 
we define $|\mu|=\mu_1+\cdots+\mu_n$ and 
$T_{x}^{\mu\delta}=T_{x_1}^{\mu_1\delta}\cdots T_{x_n}^{\mu_n\delta}$.

\begin{theorem}\label{thm:main} 
The linear difference operators $H_l=H_{l}^{x}$ $(l=0,1,2,\ldots)$ 
defined as above belong to the commutative algebra 
$\mathcal{R}=\mathbb{C}[D_1,\ldots,D_n]$ 
of Ruijsenaars operators.  In particular, one has 
\begin{equation}
D_rH_s=H_sD_r,\qquad H_r H_s=H_s H_r\qquad(r,s=0,1,2,\ldots). 
\end{equation}
\end{theorem}

We first remark that the family of difference operators $H_l$ ($l=0,1,2,\ldots$)
originates from the kernel identities for Ruijsenaars operators
% Komori-Noumi-Shiraishi 
(\cite{KNS2009}) and the 
duality transformation formulas for multiple elliptic hypergeometric 
series 
% Kajihara-Noumi and Rosengren
(\cite{KN2003}, \cite{Ro2006}).  
Let $G(z)$ be a nonzero meromorphic function on $\mathbb{C}$ 
such that  
$G(z+\delta)=[z]\,G(z)$,  
and define the kernel function $\Phi(x;y)$ of Cauchy type by
\begin{align}\label{eq:kerCauchy}
\Phi(x;y)=\dprod{i=1}{n}\dprod{k=1}{n}
\dfrac{G(x_i+y_k)}{G(x_i+y_k+\kappa)}
\end{align}
for $x=(x_1,\ldots,x_n)$ and $y=(y_1,\ldots,y_n)$.  
Then from \cite{KN2003}, Theorem 1.3 we have the kernel identity 
\begin{align}
D_{r}^{x}\,\Phi(x;y)=
D_{r}^{y}\,\Phi(x;y)
\qquad(r=0,1,\ldots,n)
\end{align}
for the Ruijsenaars operators $D_r$ ($r=0,1,\ldots,r$).  
On the other hand, 
the duality transformation formula 
for multiple elliptic hypergeometric series 
implies 
\begin{align}
&
\dsum{\mu\in\mathbb{N}^m;\,|\mu|=r}{}
\dfrac{\Delta(x+\mu\delta)}{\Delta(x)}
\dprod{i,j=1}{n}
\dfrac{[x_i-x_j+\kappa]_{\mu_i}}{[x_i-x_j+\delta]_{\mu_i}}
\dprod{i,k=1}{n}
\dfrac{[x_i+y_k]_{\mu_i}}{[x_i+y_k+\kappa]_{\mu_i}}
\nonumber\\
&=
\dsum{\nu\in\mathbb{N}^m;\,|\nu|=r}{}
\dfrac{\Delta(y+\nu\delta)}{\Delta(y)}
\dprod{k,l=1}{n}
\dfrac{[y_k-y_l+\kappa]_{\nu_k}}{[y_k-y_l+\delta]_{\nu_k}}
\dprod{k,i=1}{n}
\dfrac{[y_k+x_i]_{\nu_k}}{[y_k+x_i+\kappa]_{\nu_k}}
\quad(r=0,1,2,\ldots), 
\end{align}
as the special case where $m=n$ and $a_i=b_i=\kappa$ ($i=1,\ldots,n$) 
in the notation of \cite{KN2003}, Theorem 2.2. 
This means that 
\begin{align}
H_r^{x} \,\Phi(x;y)=H_r^{y}\, \Phi(x;y)\quad(r=0,1,2,\ldots).  
\end{align}
Namely, the kernel function for the Ruijsenaars 
operators $D_r$ ($r=0,1,\ldots,n$) simultaneously 
intertwines the operators $H_r$ ($r=0,1,2\ldots$).   
In view of this fact, it would be reasonable to expect 
that the operators $H_r$ should already belong 
to the commutative algebra $\mathbb{C}[D_1,\ldots,D_n]$ 
of Ruijsenaars operators. 
Theorem \ref{thm:main} mentioned above ensures that 
it is actually the case.  

\par\medskip
In this paper we prove 
Theorem \ref{thm:main} as a consequence of the following 
recurrence formula {\em of Wronski type} for $H_l$
($l=0,1,2,\ldots$).  
\begin{theorem}\label{thm:Wronski}
The difference operators $H_l$ $(l=0,1,2,\ldots)$ satisfy the following 
recurrence formula in relation to $D_r$ $(r=1,\ldots,n)$\,$:$ 
\begin{equation}\label{eq:Wronski}
\dsum{r+s=l}{}\,(-1)^r[r\kappa+s\delta]\,D_r\,H_s=0\qquad
(l=1,2,\ldots). 
\end{equation}
\end{theorem}

Recall that the elementary symmetric functions 
\begin{equation}
e_r=e_r(\xi)=\dsum{1\le i_1<\cdots<i_r\le m}{}\,
\xi_{i_1}\cdots \xi_{i_r}
\qquad(r=0,1,2,\ldots)
\end{equation}
in $\xi=(\xi_1,\ldots,\xi_n)$, and 
the complete homogeneous symmetric functions 
\begin{equation}
h_l=h_l(\xi)=\dsum{\mu_1+\cdots+\mu_n=l}{}\xi_1^{\mu_1}\cdots \xi_{n}^{\mu_n}
\qquad(l=0,1,2,\ldots)
\end{equation}
are related to each other through the {\em Wronski formula}
\begin{equation}
\dsum{r+s=l}{}(-1)^r e_r\,h_s=0\qquad(l=1,2,\ldots). 
\end{equation}
(See \cite{Ma1995}.)  
Theorem \ref{thm:Wronski} can be thought of as an operator version 
of this Wronski formula for symmetric functions. 
 %in the commutative algebra of Ruijsenaars operators. 
A proof of Theorem \ref{thm:Wronski} will be given 
in Section 2 by using a functional identity for $[z]$ 
(Lemma \ref{lem:keyidentity}). 

\par\medskip
From the recurrence formulas 
\begin{equation}
\begin{array}{ll}
[\delta]H_1-[\kappa]D_1=0,\\[4pt]
[2\delta]H_2-[\kappa+\delta]D_1H_1+[2\kappa]D_2=0,\\[4pt]
[3\delta]H_3-
[\kappa+2\delta]D_1H_2+[2\kappa+\delta]D_2H_1+[3\kappa]D_3=0,\\[4pt]
\qquad\ldots, 
\end{array}
\end{equation}
we see inductively that $H_l$ belongs to the commutative algebra 
$\mathcal{R}=\mathbb{C}[D_1,\ldots,D_n]$ 
of Ruijsenaars operators for all $l=0,1,2,\ldots$.  
In fact we have
\begin{equation}
\begin{array}{ll}
H_1=\dfrac{[\kappa]}{[\delta]}D_1,\\[10pt]
H_2=\dfrac{[\kappa][\kappa+\delta]}{[\delta][2\delta]}D_1^2
-\dfrac{[2\kappa]}{[2\delta]}D_2,\\[10pt]
H_3=\dfrac{[\kappa][\kappa+\delta][\kappa+2\delta]}
{[\delta][2\delta][3\delta]}D_1^3
-\dfrac{[2\kappa][\kappa+\delta]}
{[2\delta][3\delta]}D_2D_1
%\\[10pt]
%\qquad\quad
%\mbox{}
-
\dfrac{[\kappa][2\kappa+\delta]}
{[\delta][3\delta]}D_1D_2
+\dfrac{[3\kappa]}{[3\delta]}D_3,
\\[10pt]
\qquad\ldots. 
\end{array}
\end{equation}
The relationship between the two families of difference operators 
$D_r$ ($r=01,2,\ldots$) and $H_l$ ($l=0,1,2,\ldots$) is  
described as follows.

\begin{proposition}\label{prop:A}
For each $l=0,1,2,\ldots$, 
the difference operator $H_l$ is expressed in terms 
of $D_r$ $(r=0,1,\ldots)$ by the determinant formula 
\begin{equation}
H_l=\det\left(\dfrac{[(i\!-\!j\!+\!1)\kappa\!+\!(j\!-\!1)\delta]}
{[i\delta]}D_{i-j+1}\right)_{i,j=1}^{l}\qquad(l=0,1,2,\ldots). 
\end{equation}
Conversely,
\begin{equation}
D_l=\det\left(\dfrac{[(i\!-\!j\!+\!1)\delta\!+\!(j\!-\!1)\kappa]}
{[i\kappa]}H_{i-j+1}\right)_{i,j=1}^{l}\qquad(l=0,1,2,\ldots). 
\end{equation}
\end{proposition}

\begin{proposition}\label{prop:B}
For each $l=1,2,\ldots$, 
the difference operator $H_l$ is expressed explicitly as 
\begin{equation}
H_l=\dsum{d=1}{l}\ 
(-1)^{l-d}
\dsum{r_1+\cdots+r_d=l; r_i\ge 1}{}\,
\left(\dprod{i=1}{d}
\dfrac{[(r_1\!+\!\cdots\!+\!r_{i-1})\delta+r_i\kappa]}
{[(r_1+\cdots+r_i)\delta]}\right)\,D_{r_1}\cdots D_{r_d}
\end{equation}
in terms of $D_r$ $(r=0,1,\ldots)$. 
\end{proposition}

We also summarize the kernel identities relevant to 
the difference operators $D_r$ and $H_r$ for the sake of reference. 
\begin{theorem}
$(1)$ \ \ For two sets of variables $x=(x_1,\ldots,x_n)$ and $y=(y_1,\ldots,y_n)$, 
the kernel function $\Phi(x,y)$ of Cauchy type in \eqref{eq:kerCauchy} satisfies 
the following two types of kernel identities $:$
\begin{align}
(DD)\qquad&\quad 
D^x_r\Phi(x;y)=D^y_r\Phi(x;y)\qquad(r=0,1,\ldots,n),
\\[4pt]
(HH)\qquad&\quad H^x_r\Phi(x;y)=H^y_r\Phi(x;y)\qquad(r=0,1,2,\ldots). 
\end{align}
$(2)$\ \ For two sets of variables $x=(x_1,\ldots,x_m)$ and $y=(y_1,\ldots,y_n)$, 
let 
\begin{equation}
\Psi(x;y)=\prod_{i=1}^{m}\prod_{k=1}^{n} [x_i-y_k]
\end{equation}
be the kernel function of dual Cauchy type.  
Under the balancing condition $m\kappa+n\delta=0$, 
$\Psi(x;y)$ satisfies the kernel identity 
\begin{equation}
(HD)\qquad\quad
H^x_r \Psi(x;y)=(-1)^r\widehat{D}^y_r \Psi(x;y)\qquad(r=0,1,2,\ldots),
\end{equation}
where $\widehat{D}^y_r$ denotes the difference operator obtained 
from $D^y_r$ by exchanging the parameters $\delta$ and $\kappa$.  
\end{theorem}

Propositions \ref{prop:A} and \ref{prop:B}  
are consequences of the recurrence 
formula of Wronski type (see Section 3). 
After a complement on kernel identities 
for the Ruijsenaars operators (Section 4), 
we finally give some remarks on the trigonometric case in Section 5. 

\begin{quote}
{\bf Notes:}\ \ This paper is based on 
a collaboration of the authors which was completed as 
master's thesis \cite{Sano2008} of the second author 
in Japanese.  Also, an earlier version of the present paper, 
written around 2012, has been circulated among some researchers.  
For these reasons, some of the results in this paper are already cited 
in several literatures \cite{FHHSY2009, NS2012, Masuda2013, BCR2015}
with reference to a private communication or to 
a paper in preparation.  
\end{quote}

\section{Recurrence formula of Wronski type}
%%%%%%%%%%%%%%%%%%%%%%%%%%%%%%%%%%%%%%%%%%%%%%%
In this section we give a proof of Theorem \ref{thm:Wronski}.  
Our goal is to establish the recurrence formula
of Wronski type between the two sequences of difference 
operators $D_r$ ($r=0,1,2,\ldots$) and
$H_s$ ($s=0,1,2,\ldots$).  

\begin{theorem}\label{thm:WronskiS2}
The difference operators $H_l$ $(l=0,1,2\ldots)$ 
defined by \eqref{eq:defH} satisfy the 
recurrence formulas
\begin{equation}\label{eq:WR}
\dsum{r+s=l}{}(-1)^r [r\kappa+s\delta]\,D_r H_s=0\qquad(l=1,2,\ldots). 
\end{equation}
\end{theorem}
Since $D_0=1$, by this theorem 
we see inductively that $H_l$ belong to 
$\mathbb{C}[D_1,\ldots,D_n]$ for all $l=0,1,2,\ldots$. 
\begin{theorem}
The difference operators $H_l$ $(l=0,1,2\ldots)$ belong 
to the commutative algebra $\mathbb{C}[D_1,\ldots,D_n]$ 
of Ruijsenaars operators.  In particular one has 
\begin{equation}
D_rH_s=H_sD_r,\qquad H_r H_s=H_s H_r\qquad(r,s=0,1,2,\ldots). 
\end{equation}
\end{theorem}

\par\medskip
\proof{of Theorem \ref{thm:WronskiS2}}
We express the difference operators $D_r$ as 
\begin{equation}
D_{r}=\dsum{|I|=r}{}A_I(x) T_{x}^{\,\epsilon_I\delta},
\qquad
A_I(x)=\dprod{i\in I;\,j\notin I}{}\dfrac{[x_i-x_j+\kappa]}{[x_i-x_j]}
=\dfrac{\Delta(x+\epsilon_I\kappa)}{\Delta(x)}
\end{equation}
where we define $\epsilon_I=\dsum{i\in I}{}{\epsilon_i}$ by using the unit vectors  
$\epsilon_1,\ldots,\epsilon_n$ of $\mathbb{N}^n$. 
Similarly we express $H_l$ as 
\begin{equation}
H_=\dsum{|\mu|=l}{} H_\mu(x)T_{x}^{\mu\delta},
\quad
H_{\mu}=\dfrac{\Delta(x+\mu\delta)}{\Delta(x)}\,\dprod{i,j=1}{n}
\dfrac{[x_j-x_i+\kappa]_{\mu_j}}{[x_j-x_i+\delta]_{\mu_j}}. 
\end{equation}
When $r+s=l$, we compute
\begin{equation}
\arraycolsep=2pt
\begin{array}{ll}
D_rH_s&=\dsum{|I|=r}{}\ \dsum{|\mu|=s}{}A_I(x) H_\mu(x+\epsilon_I\delta)T_{x}^{(\epsilon_I+\mu)\delta}
\\
&=\dsum{|\lambda|=l}{}\ 
\dsum{I\subseteq \mbox{\scriptsize supp}(\lambda);\,|I|=r}{}\,
A_I(x) H_{\lambda-\epsilon_I}(x+\epsilon_I\delta)T_{x}^{\lambda\delta},
\end{array}
\end{equation}
where $\mbox{supp}(\lambda)=\{i\in\{1,\ldots,n\}\ |\ \lambda_i>0\}$. 
Hence, the recurrence formula \eqref{eq:WR} is equivalent to saying that 
\begin{equation}\label{eq:WR1}
\dsum{I\subseteq\mbox{\scriptsize supp}(\lambda)}{}
(-1)^{|I|}\big[\,|I|\kappa+(|\lambda|-|I|)\delta\,\big]\, 
A_I(x) H_{\lambda-\epsilon_I}(x+\epsilon_I\delta)=0
\end{equation}
for any $\lambda\in\mathbb{N}^n$ with $|\lambda|>0$. 
We now make the expression $A_I(x)H_{\lambda-\epsilon_I}(x+\epsilon_I)$ explicit. 
Setting $L=\mbox{supp}(\lambda)$, we have
\begin{equation}
\arraycolsep=2pt
\begin{array}{ll}
A_I(x)H_{\lambda-\epsilon_I}(x+\epsilon_I)
&=
\dfrac{\Delta(x+\epsilon_I\kappa)}{\Delta(x)}
\dfrac{\Delta(x+\lambda\delta)}{\Delta(x+\epsilon_I\delta)}
\\[10pt]
&\quad\cdot
\dprod{i\in I; j\in I}{}
\dfrac{[x_j-x_i+\kappa]_{\lambda_j-1}}{[x_j-x_i+\delta]_{\lambda_j-1}}
\dprod{i\in I; j\in L\backslash I}{}
\dfrac{[x_j-x_i+\kappa-\delta]_{\lambda_j}}{[x_j-x_i]_{\lambda_j}}
\\
&\quad\cdot
\dprod{i\notin I; j\in I}{}
\dfrac{[x_j-x_i+\kappa+\delta]_{\lambda_j-1}}{[x_j-x_i+2\delta]_{\lambda_j-1}}
\dprod{i\notin I; j\in L\backslash I}{}
\dfrac{[x_j-x_i+\kappa]_{\lambda_j}}{[x_j-x_i+\delta]_{\lambda_j}}. 
\end{array}
\end{equation}
Noting that 
\begin{equation}
\dfrac{\Delta(x+\epsilon_I\kappa)}{\Delta(x)}
\dfrac{\Delta(x+\lambda\delta)}{\Delta(x+\epsilon_I\delta)}
=
\dfrac{\Delta(x+\lambda\delta)}{\Delta(x)}
\dfrac{\Delta(x+\epsilon_I\kappa)}{\Delta(x+\epsilon_I\delta)}
=
\dfrac{\Delta(x+\lambda\delta)}{\Delta(x)}
\dprod{i\notin I; j\in I}{}
\dfrac{[x_j-x_i+\kappa]}{[x_j-x_i+\delta]}, 
\end{equation}
we can compute $A_I(x)H_{\lambda-\epsilon_I}(x+\epsilon_I)$ as follows: 
\begin{align}
A_I(x)H_{\lambda-\epsilon_I}(x+\epsilon_I)
&=
\dfrac{\Delta(x+\lambda\delta)}{\Delta(x)}
\dprod{i\notin I; j\in L}{}
\dfrac{[x_j-x_i+\kappa]_{\lambda_j}}{[x_j-x_i+\delta]_{\lambda_j}}
\nonumber\\
&\qquad\cdot
\dprod{i\in I; j\in I}{}
\dfrac{[x_j-x_i+\kappa]_{\lambda_j-1}}{[x_j-x_i+\delta]_{\lambda_j-1}}
\dprod{i\in I; j\in L\backslash I}{}
\dfrac{[x_j-x_i+\kappa-\delta]_{\lambda_j}}{[x_j-x_i]_{\lambda_j}}
\nonumber\\
&=
\dfrac{\Delta(x+\lambda\delta)}{\Delta(x)}
\dprod{i\in\{1,\ldots,n\}; j\in L}{}
\dfrac{[x_j-x_i+\kappa]_{\lambda_j}}{[x_j-x_i+\delta]_{\lambda_j}}
\nonumber\\
&\qquad\cdot
\dprod{i\in I; j\in L\backslash I}{}
\dfrac{[x_j\!-\!x_i\!+\!\kappa\!-\!\delta]}{[x_j\!-\!x_i]}. 
\dprod{i\in I; j\in L}{}
\dfrac{[x_j\!-\!x_i\!+\!\lambda_i\delta]}
{[x_j\!-\!x_i\!+\!\kappa\!+\!(\lambda_j\!-\!1)\delta]}. 
\end{align}
Hence \eqref{eq:WR1} is equivalent to
\begin{equation}
\begin{array}{ll}
\dsum{I\subseteq L}{}
(-1)^{|I|}\big[\,|I|\kappa\!+\!(|\lambda|\!-\!|I|)\delta\,\big]\, 
%\\
%\quad\cdot
\dprod{i\in I; j\in L\backslash I}{}
\dfrac{[x_j\!-\!x_i\!+\!\kappa\!-\!\delta]}{[x_j\!-\!x_i]}
\dprod{i\in I; j\in L}{}
\dfrac{[x_j\!-\!x_i\!+\!\lambda_j\delta]}
{[x_j\!-\!x_i\!+\!\kappa\!+\!(\lambda_j\!-\!1)\delta]}
=0
\end{array}
\end{equation}
for any $L\ne\phi$ and $\lambda\in\mathbb{N}^n$ with $\mbox{supp}(\lambda)=L$. 
Setting $\lambda=\epsilon_L+\nu$, 
we rewrite this in the form 
\begin{equation}
\begin{array}{ll}
\dsum{I\subseteq L}{}
(-1)^{|I|}\big[\,|I|\kappa\!+\!(|\nu|\!+\!|L|\!-\!|I|)\delta\,\big]\, 
%\\
%\quad\cdot
\dprod{i\in I; j\in L\backslash I}{}
\dfrac{[x_j\!-\!x_i\!+\!\kappa\!-\!\delta]}{[x_j\!-\!x_i]}
\dprod{i\in I; j\in L}{}
\dfrac{[x_j\!-\!x_i\!+\!\delta\!+\!\nu_j\delta]}
{[x_j\!-\!x_i\!+\!\kappa\!+\nu_j\delta]}
=0
\end{array}
\end{equation}
for any $\nu\in\mathbb{N}^n$ with $\mbox{supp}(\nu)\subseteq L$.
Since this formula contains only those variables $x_i$ with $i\in L$, we 
have only to consider the case where $L=\{1,\ldots,n\}$ ($n\ge1$): 
\begin{equation}
\begin{array}{ll}
\dsum{I\subseteq\{1,\ldots,n\}}{}
(-1)^{|I|}\big[\,|I|\kappa\!+\!(|\nu|\!+\!m\!-\!|I|)\delta\,\big]
\\
\qquad\cdot
\dprod{i\in I; j\notin I}{}
\dfrac{[x_j\!-\!x_i\!+\!\kappa\!-\!\delta]}{[x_j\!-\!x_i]}
\dprod{i\in I; j\in\{1,\ldots,n\}}{}
\dfrac{[x_j\!-\!x_i\!+\!\delta\!+\!\nu_j\delta]}
{[x_j\!-\!x_i\!+\!\kappa\!+\nu_j\delta]}=0
\end{array}
\end{equation}
for any $\nu\in\mathbb{N}^n$. 
This identity follows from the following functional identity 
by the change of variables 
\begin{equation}
z_i=x_i,\quad w_i=x_i+\delta+\nu_i\delta\quad(i=1,\ldots,n);\quad
a=\kappa-\delta. 
\end{equation}

\begin{lemma}\label{lem:keyidentity}
Given two sets of variables $z=(z_1,\ldots,z_n)$, $w=(w_1,\ldots,w_n)$
and a parameter $a$, the following identity holds as a meromorphic function 
in $(z_1,\ldots,z_n,w_1,\ldots,w_n)$ for $n\ge1$$:$ 
\begin{equation}\label{eq:keyidentity}
\dsum{I\subseteq\{1,\ldots,n\}}{}
(-1)^{|I|}\dfrac{[|w|-|z|+|I|a]}{[\,|w|-|z|\,]}
\dprod{i\in I; j\notin I}{}
\dfrac{[z_j\!-\!z_i\!+a]}{[z_j\!-\!z_i]}
\dprod{i\in I;\,k\in\{1,\ldots,n\}}{}
\dfrac{[w_k-z_i]}
{[w_k-z_i+a]}=0, 
\end{equation}
where $|z|=\sum_{i=1}^nz_i$, $|w|=\sum_{k=1}^nw_k$. 
\end{lemma}

\proof{}
We give a proof of the functional identity \eqref{eq:keyidentity} for 
the case where $[z]=\sigma(z;\Omega)$ is 
the Weierstrass sigma function associated with a period lattice $\Omega$. 
By the classification of $[z]$, 
it is not difficult to derive 
\eqref{eq:keyidentity} 
for any $[z]$ in this class from the 
case of $\sigma(z;\Omega)$,  
by the limiting procedures from $\sigma(z;\Omega)$ to 
$\sin(\pi z/\omega)$ and $z$, and by the 
invariance of \eqref{eq:keyidentity} under the transformation 
$[z] \to e^{cz^2}[z]$. 

Identity \eqref{eq:keyidentity} for $[z]=\sigma(z;\Omega)$ can be 
proved by the induction on $n$. 
Since it holds trivially for $n=1$, we assume $n\ge2$. 
We regard the left-hand side of \eqref{eq:keyidentity} 
as a meromorphic function of $w_n$, and 
denote it by $F(w_n)$ assuming that the other variables are generic. 
Note first that 
$F(w_n)$ is an elliptic function possibly with simple poles 
at $w_n\equiv z_1-a,\ldots, z_n-a$ and $w_n\equiv |z|-|w'|$ modulo 
the period lattice $\Omega$, 
where $w'=(w_1,\ldots,w_{n-1})$. 
We first compute the residue of $F(w_n)$ at $w_n=z_n-a$.  
Nontrivial residues possibly arise from the terms corresponding 
to $I$ containing $n$; we parametrize such $I$'s as 
$I=J\cup\{n\}$ with $J\subseteq\{1,\ldots,n-1\}$. 
Then we have 
\begin{equation}
\arraycolsep=2pt
\begin{array}{l}
\mbox{Res}_{w_n=z_n-a}(F(w_n)dw_n)\\[8pt]
=
\dfrac{[a]
[|w'|-|z'|]}{[|w'|\!-\!|z'|\!-\!a]}
\dprod{j\in\{1,\ldots,n-1\}}{}
\dfrac{[z_j-z_n+a]}{[z_j-z_n]}
\dprod{k\in\{1,\ldots,n-1\}}{}
\dfrac{[w_k-z_n]}
{[w_k-z_n+a]}
\\[16pt]
\quad\cdot
\dsum{J\subseteq\{1,\ldots,n-1\}}{}
(-1)^{|J|}\dfrac{[|w'|\!-\!|z'|\!+\!|J|a]}{[|w'|-|z'|]}
\\
\qquad\qquad\cdot
\dprod{i\in J; j\in\{1,\ldots,n-1\}\backslash J}{}
\dfrac{[z_j\!-\!z_i\!+a]}{[z_j\!-\!z_i]}
\dprod{i\in J; k\in\{1,\ldots,n-1\}}{}
\dfrac{[w_k-z_i]}
{[w_l\!-\!z_i\!+\!a]}\\[16pt]
=0 
\end{array}
\end{equation}
by the induction hypothesis ($z'=(z_1,\ldots,z_{n-1})$). 
Since $F(w_n)$ is symmetric with respect to 
$(z_1,\ldots,z_n)$, we see that 
$w_n\equiv z_1-a,\ldots,z_n-a$ are all removable poles of $F(w_n)$.
Hence $F(w_n)$ has at most one simple pole in each fundamental parallelogram, 
which is impossible unless $F(w_n)$ is a constant function since 
it is an elliptic function. 
We next look at the value of $F(w_n)$ at $w_n=z_n$. 
It is computed as 
\begin{equation}
\arraycolsep=2pt
\begin{array}{ll}
F(z_n)&=
\dsum{I\subseteq\{1,\ldots,n-1\}}{}
(-1)^{|I|}\dfrac{[|w'|-|z'|+|I|a]}{[|w'|-|z'|]}
\\
&\qquad\cdot
\dprod{i\in\{1,\ldots,n-1\}\backslash I; j\in I}{}
\dfrac{[z_j\!-\!z_i\!+a]}{[z_j\!-\!z_i]}
\dprod{i\in I; k\in\{1,\ldots,n-1\}}{}
\dfrac{[w_k-z_i]}
{[w_k-z_i+a]}
\\
&=0
\end{array}
\end{equation}
again by the induction hypothesis.  This implies that $F(w_n)$ is 
identically zero as a meromorphic function of $w_n$. 
\qed

%We give some remarks on the functional identity \eqref{eq:keyidentity}. 

\begin{remark}\rm 
Lemma \ref{lem:keyidentity} can be proved in a different way if we make 
use of the argument of \cite{KN2003}. 
%As in \cite{KN2003}, let us denote by $[z]$, in place of $\sigma(z)$ above, 
%any nonzero entire function 
%satisfying the Riemann relation. 
Recall that in (1.14) of Section 1, \cite{KN2003}, 
the following identity is derived from the determinant formula of Frobenius: 
\begin{equation}
\arraycolsep=2pt
\begin{array}{ll}
\dfrac{E(T_z;u)D(z;w)}{D(z;w)}\\
=\dsum{I\subseteq\{1,\ldots,n\}}{}
u^{|I|}\dfrac{[\lambda+|z|+|w|+|I|\delta]}{[\lambda+|z|+|w|]}
\dprod{i\in I; j\notin I}{}\dfrac{[z_i-z_j+\delta]}{[z_i-z_j]}
\dprod{i\in I; j\in\{1,\ldots,n\}}{}
\dfrac{[z_i+w_k]}{[z_i+w_k+\delta]},
\end{array}
\end{equation}
where $D(z;w)$ is the Frobenius determinant 
\begin{equation}
D(z;w)=
\det\left(\dfrac{[\lambda+z_i+w_j]}{[\lambda][z_i+w_j]}\right)_{i,j=1}^{n}
=
\dfrac{[\lambda+|z|+|w|]\Delta(z)\Delta(w)}{[\lambda]\ \tprod{i,j=1}{n}[z_i+w_j]},
%\quad
%\Delta(z)=\dprod{1\le i<j\le n}{}[z_i-z_j]
\end{equation}
and $E(T_z;u)=\tprod{i=1}{n}(1+uT_{z_i}^{\delta})$. 
This means that 
\begin{equation}
\arraycolsep=2pt
\begin{array}{ll}
\dfrac{[\lambda]\ \tprod{i,j=1}{n}[z_i+w_j]}
{[\lambda+|z|+|w|]\Delta(z)\Delta(w)}
\det\left(
\dfrac{[\lambda+z_i+w_j]}{[\lambda][z_i+w_j]}
+u\dfrac{[\lambda+z_i+w_j+\delta]}{[\lambda][z_i+w_j+\delta]}
\right)_{i,j=1}^{n}
\\[16pt]
=
\dsum{I\subseteq\{1,\ldots,n\}}{}
u^{|I|}\dfrac{[\lambda+|z|+|w|+|I|\delta]}{[\lambda+|z|+|w|]}
\dprod{i\in I; j\notin I}{}\dfrac{[z_i-z_j+\delta]}{[z_i-z_j]}
\dprod{i\in I; j\in\{1,\ldots,n\}}{}
\dfrac{[z_i+w_k]}{[z_i+w_k+\delta]}. 
\end{array}
\end{equation}
By setting $u=-1$, we obtain
\begin{equation}
\arraycolsep=2pt
\begin{array}{ll}
\dfrac{[\lambda]\ \tprod{i,j=1}{n}[z_i+w_j]}
{[\lambda+|z|+|w|]\Delta(z)\Delta(w)}
\det\left(
\dfrac{1}{[\lambda]}
\left(\dfrac{
[\lambda+z_i+w_j]}{[z_i+w_j]}
-\dfrac{[\lambda+z_i+w_j+\delta]}{[z_i+w_j+\delta]}
\right)
\right)_{i,j=1}^{n}
\\[16pt]
=
\dsum{I\subseteq\{1,\ldots,n\}}{}
(-1)^{|I|}\dfrac{[\lambda+|z|+|w|+|I|\delta]}{[\lambda+|z|+|w|]}
\dprod{i\in I; j\notin I}{}\dfrac{[z_i-z_j+\delta]}{[z_i-z_j]}
\dprod{i\in I; j\in\{1,\ldots,n\}}{}
\dfrac{[z_i+w_k]}{[z_i+w_k+\delta]}. 
\end{array}
\end{equation}
Note that in the limit 
$\lambda\to 0$, 
each entry of the matrix of the left-hand side 
has a finite limit
\begin{equation}
\lim_{\lambda\to0}
\dfrac{1}{[\lambda]}
\left(
\dfrac{
[\lambda+z_i+w_j]}{[z_i+w_j]}
-\dfrac{[\lambda+z_i+w_j+\delta]}{[z_i+w_j+\delta]}
\right)
=
\dfrac{[z_i+w_j]'}{[z_i+w_j]}
-\dfrac{[z_i+w_j+\delta]'}{[z_i+w_j+\delta]}. 
\end{equation}
Hence the left-hand side converges to zero as $\lambda\to 0$.  This implies that 
\begin{equation}
\dsum{I\subseteq\{1,\ldots,n\}}{}
(-1)^{|I|}\dfrac{[|z|+|w|+|I|\delta]}{[|z|+|w|]}
\dprod{i\in I; j\notin I}{}\dfrac{[z_i-z_j+\delta]}{[z_i-z_j]}
\dprod{i\in I; j\in\{1,\ldots,n\}}{}
\dfrac{[z_i+w_k]}{[z_i+w_k+\delta]}
=0
\end{equation}
By replacing each $z_i$ with $-z_i$, and $\delta$ with $a$, 
we obtain Lemma \ref{lem:keyidentity}.
\end{remark}

\section{Explicit relations between the two commuting families}

Setting
\begin{align}
D_r^{(l)} = \dfrac{[r\kappa+(l-r)\delta]}{[l\delta]}D_r\qquad(0\le r\le l), 
\end{align}
we rewrite the recurrence formula of Theorem \ref{thm:WronskiS2}
as
\begin{align}
(-1)^l H_l+(-1)^{l-1}D_1^{(l)}H_{l-1}+\cdots-D_{l-1}^{(l)}H_1=-D_l^{(l)}\qquad(l=1,2,\ldots). 
\end{align}
In the matrix form this means that 
\begin{align}
\left[
\begin{matrix}
1 & 0 & 0 & \ldots &\ 0\ \\
D_1^{(2)} & 1 & 0&\ldots &\vdots\\
D_2^{(3)} & D_1^{(3)} & 1 &\\
\vdots & \vdots &&\ddots & 0 \\
D_{l-1}^{(l)} & D_{l-2}^{(l)} & \ldots & D_{1}^{(l)} & 1
\end{matrix}
\right]
\left[
\begin{matrix}
-H_1 \\[3pt] 
H_2 \\[3pt] 
-H_3 \\[3pt] 
\vdots \\[3pt] 
(-1)^l H_l
\end{matrix}
\right]
=
-\left[
\begin{matrix}
\ \ D_1^{(1)}\ \ \\[1pt]
D_2^{(2)} \\[1pt] 
D_3^{(3)} \\[1pt] 
\vdots \\[1pt]
D_l^{(l)}
\end{matrix}
\right].
\end{align}
Hence by Cramer's formula we obtain
\begin{align}
H_l=\det
\left[\ 
\begin{matrix}
D_1^{(1)}& 1 & 0 & \ldots &\ 0\ \\
D_2^{(2)}&D_1^{(2)} & 1& &\vdots\\
\vdots &D_2^{(3)} &D_1^{(3)} &\ddots &\vdots\\
\vdots& \vdots &  \ddots& \ddots&1 \\
D_l^{(1)}&D_{l-1}^{(l)} & \ldots& \ldots & D_{1}^{(l)}
\end{matrix}
\right]
=
\det\left(D_{i-j+1}^{(i)}\right)_{i,j=1}^{l}
\qquad(l=1,2,\ldots).  
\end{align}
Namely, we have 
\begin{align}
H_l
=
\det\left(\dfrac{[(i-j+1)\kappa+(j-1)\delta]}{[i\delta]}D_{i-j+1}\right)_{i,j=1}^{l}
\qquad(l=1,2.\ldots). 
\end{align}
By symmetry, we also have 
\begin{align}
D_l
=
\det\left(\dfrac{[(i-j+1)\delta+(j-1)\kappa]}{[i\kappa]}H_{i-j+1}\right)_{i,j=1}^{l}
\qquad(l=1,2.\ldots).
\end{align}

The recurrence formula \eqref{eq:WR} can also be written as
\begin{align}
H_l=D_1^{(l)}H_{l-1}-\cdots+(-1)^{l-2}D_{l-1}^{(l)}H_1+(-1)^{l-1}D_{l}^{(l)}
\qquad(l=1,2,\ldots). 
\end{align}
Applying this formula repeatedly, we obtain
\begin{align}
H_l
&=
\dsum{d=1}{l}\ 
(-1)^{l-d}
\dsum{0=l_0<l_1<\cdots<l_d=l}{}
D_{l_d-l_{d-1}}^{(l_d)}D_{l_{d-1}-l_{d-2}}^{(l_{d-1})}\cdots D_{l_{1}-l_{0}}^{(l_{1})}
\nonumber\\
&=
\dsum{d=1}{l}\ 
(-1)^{l-d}
\dsum{r_1+\cdots+r_d=l; r_i>0}{}\ 
\dprod{i=1}{d}
D_{r_i}^{(r_1+\cdots+r_i)}
\nonumber\\
&=
\dsum{d=1}{l}\ 
(-1)^{l-d}
\dsum{r_1+\cdots+r_d=l; r_i>0}{}\ 
\left(\dprod{i=1}{d}
\dfrac{[(r_1+\cdots+r_{i-1})\delta+r_i\kappa]}
{[(r_1+\cdots+r_i)\delta]}
\right)
D_{r_1}\cdots D_{r_d}. 
\end{align}

\section{Kernel identities}
We recall from \cite{KN2003} the duality transformation formula for 
multiple elliptic hypergeometric series (of type $A$):\ 
Under the balancing condition $a_1+\cdots+a_m=b_1+\cdots+b_n$, 
\begin{align}
&
\dsum{\mu\in\mathbb{N}^m;\, |\mu|=r}{}
\dfrac{\Delta(x+\mu\delta)}{\Delta(x)}
\dprod{i,j=1}{m}
\dfrac{[x_i-x_j+a_j]_{\mu_i}}{[x_i-x_j+\delta]_{\mu_i}}
\dprod{i=1}{m}
\dprod{k=1}{n}
\dfrac{[x_i+y_k-b_k]_{\mu_i}}{[x_i+y_k]_{\mu_i}}
\nonumber\\
&=
\dsum{\nu\in\mathbb{N}^n;\, |\nu|=r}{}
\dfrac{\Delta(y+\nu\delta)}{\Delta(y)}
\dprod{k,l=1}{n}
\dfrac{[y_k-y_l+b_l]_{\nu_k}}{[y_k-y_l+\delta]_{\nu_k}}
\dprod{k=1}{n}
\dprod{i=1}{m}
\dfrac{[y_k+x_i-a_i]_{\nu_k}}{[y_k+x_i]_{\nu_k}}
\qquad(r=0,1,2,\ldots), 
\end{align}
where $x=(x_1,\ldots,x_m)$ and $y=(y_1,\ldots,y_n)$. 
As we already remarked, 
when $m=n$ and $a_i=b_i=\kappa$ ($i=1,\ldots,n$) this implies
\begin{align}
H_r^x \Phi(x;y)=H_r^y \Phi(x;y)\qquad(r=0,1,2,\ldots).  
\end{align}

Let $a_1=\cdots=a_m=\kappa$, 
$b_1=\cdots=b_n=-\delta$.   Then this transformation formula implies that 
under the condition $m\kappa+n\delta=0$, 
\begin{align}
&
\dsum{\mu\in\mathbb{N}^m;\, |\mu|=r}{}
\dfrac{\Delta(x+\mu\delta)}{\Delta(x)}
\dprod{i,j=1}{m}
\dfrac{[x_i-x_j+\kappa]_{\mu_i}}{[x_i-x_j+\delta]_{\mu_i}}
\dprod{i=1}{m}
\dprod{k=1}{n}
\dfrac{[x_i+y_k+\mu_i\delta]}{[x_i+y_k]}
\nonumber\\
&=
(-1)^r
\dsum{K\subseteq\pr{1,\ldots,n};\ |K|=r}{}\ 
\dprod{k\in K,\,l\notin K}{}
\dfrac{[y_k-y_l-\delta]}{[y_k-y_l]}
\dprod{k\in K}{}
\dprod{i=1}{m}
\dfrac{[y_k+x_i-k_i]}{[y_k+x_i]}
\qquad(r=0,1,2,\ldots). 
\end{align}
By replacing $y_k$ by $-y_k$ for $k=1,\ldots,n$, we obtain
\begin{align}
&
\dsum{\mu\in\mathbb{N}^m;\, |\mu|=r}{}
\dfrac{\Delta(x+\mu\delta)}{\Delta(x)}
\dprod{i,j=1}{m}
\dfrac{[x_i-x_j+\kappa]_{\mu_i}}{[x_i-x_j+\delta]_{\mu_i}}
\dprod{i=1}{m}
\dprod{k=1}{n}
\dfrac{[x_i-y_k+\mu_i\delta]}{[x_i-y_k]}
\nonumber\\
&=
(-1)^r
\dsum{K\subseteq\pr{1,\ldots,n};\ |K|=r}{}\ 
\dprod{k\in K,\,l\notin K}{}
\dfrac{[y_k-y_l+\delta]}{[y_k-y_l]}
\dprod{k\in K}{}
\dprod{i=1}{m}
\dfrac{[x_i-y_k-\kappa]}{[x_i-y_k]}
\qquad(r=0,1,2,\ldots). 
\end{align}
This means that the dual Cauchy kernel
\begin{align}
\Psi(x;y)=\dprod{i=1}{m}\dprod{k=1}{n}[x_i-y_k]
\end{align}
satisfies
\begin{align}
H_r^{(x;\delta,\kappa)} \Psi(x;y)=
(-1)^r D_r^{(y;\kappa,\delta)}\Psi(x;y)
\qquad(r=0,1,2,\ldots)
\end{align}
under the condition $m\kappa+n\delta=0$.

\section{The trigonometric cases}
In this section we consider the trigonometric cases 
where $[x]=e(x/2)-e(-x/2)$ in the notation $e(u)=\exp(2\pi\sqrt{-1}u)$ 
of the exponential function.  
In stead of the parameter $\delta$, $\kappa\in\mathbb{C}$, 
we use the multiplicative parameters $q=e(\delta)$ and $t=e(\kappa)$ 
assuming that $\Im(\delta)>0$ so that $|q|<1$.  
Note that when $z=e(x)$, we have 
$[x]=z^{\hf}-z^{-\hf}=-z^{-\hf}(1-z)$, and hence 
\begin{align}
[x]_k=(-1)^k q^{-\hf\binom{k}{2}}z^{-\hf}(z;q)_k\qquad(k=0,1,2,\ldots), 
\end{align}
in the standard notation $(z;q)_k=(1-z)(1-qz)\cdots(1-q^{k-1}z)$ of
$q$-shifted factorials.  

\par\medskip
We denote by $z=(z_1,\ldots,z_n)$ the multiplicative variables defined by 
$z_i=e(x_i)$ ($i=1,\ldots,n$) corresponding to $x=(x_1,\ldots,x_n)$.   
For these $z$ variables, 
we denote by $T_{q,z_i}$ the $q$-shift operator
with respect to $z_i$ ($i=1,\ldots,n$) and set 
$T_{q,z}^{\mu}=T_{q,z_1}^{\mu_1}\cdots T_{q,z_n}^{\mu_n}$ 
for each multi-index $\mu=(\mu_1,\ldots,\mu_n)\in\mathbb{N}^n$.  
In this multiplicative notation, it is convenient to introduce the 
$q$-difference operators $\mathcal{D}_r^{z}$
and $\mathcal{H}_l^z$ normalized so that 
\begin{align}
D_r^x=t^{-\hf r(n-r)}\mathcal{D}_r^{z}\quad(r=0,1,\ldots,n),
\quad
H_l^x=q^{-\hf l}t^{-\hf ln}\,\mathcal{H}_l^{z}\quad(l=0,1,2,\ldots). 
\end{align}
These $q$-difference operators are given explicitly by 
\begin{align}
\mathcal{D}_r^z=t^{\binom{r}{2}}\dsum{I\subseteq\pr{1,\ldots,n};\,|I|=r}{}\,
\dprod{i\in I;\,j\notin I}{}
\dfrac{tz_i-z_j}{z_i-z_j}\,T_{q,z}^{\epsilon_I}
\end{align}
and
\begin{align}
\mathcal{H}_l^z=
\dsum{\mu\in\mathbb{N}^n;\,|\mu|=l}{}\ 
\dprod{1\le i<j\le n}{}
\dfrac{q^{\mu_i}z_i-q^{\mu_j}z_j}{z_i-z_j}
\dprod{1\le i,j\le n}{}
\dfrac{(tz_i/z_j;q)_{\mu_i}}{(qz_i/z_j;q)_{\mu_i}}\,
T_{q,x}^{\mu}.  
\end{align}
The recurrence relation \eqref{eq:WR} of Wronski type 
is then rewritten as follows: 
\begin{align}\label{eq:recW}
\dsum{r+s=l}{}(-1)^r(1-t^rq^s)\mathcal{D}_r^z\, \mathcal{H}_s^z=0\quad(l=1,2,\ldots).  
\end{align}

It is known by \cite{Ma1995} that the commuting family 
of $q$-difference operators 
$\mathcal{D}_r^{z}$ ($r=0,1,\ldots,n$) 
act on the ring 
$\mathbb{C}[z]^{\mathfrak{S}_n}=\mathbb{C}[z_1,\ldots,z_n]^{\mathfrak{S}_n}$ 
of symmetric polynomials in $z=(z_1,\ldots,z_n)$, 
and that they are simultaneously diagonalized by the (monic) 
Macdonald polynomials $P_\lambda(z)=P_\lambda(z|q,t)$ indexed by 
partitions $\lambda=(\lambda_1,\ldots,\lambda_n)$ with $l(\lambda)\le n$:  
\begin{align}\label{eq:DonP}
\mathcal{D}_r^z P_{\lambda}(z)=P_{\lambda}(z)\,e_r(t^\delta q^\lambda)
\quad(r=0,1,2,\ldots,n), 
\end{align}
where $e_r(\xi)$ stands for the elementary symmetric polynomial 
of degree $r$ for each $r=0,1,\ldots,n$, and 
$t^\delta q^{\lambda}=(t^{n-1}q^{\lambda_1},t^{n-2}q^{\lambda_2},\ldots,q^{\lambda_n})$. 
In terms of the generating function 
\begin{align}
\mathcal{D}^z(u)=\dsum{r=0}{n}(-u)^r \mathcal{D}_r^{z}
=
\dsum{I\subseteq\pr{1,\ldots,n}}{}
t^{\binom{|I|}{2}}
\dprod{i\in I;\,j\notin I}{}
\dfrac{tz_i-z_j}{z_i-z_j}\,T_{q,z}^{\epsilon_I}, 
\end{align}
formula \eqref{eq:DonP} is equivalent to 
\begin{align}
\mathcal{D}^z(z) P_{\lambda}(z)=P_{\lambda}(z)\,\dprod{i=1}{n}(1-ut^{n-i}q^{\lambda_i}).  
\end{align}

Since $\mathcal{H}_l^z\in\mathbb{C}[\mathcal{D}_1^z,\ldots,\mathcal{D}_n^z]$, 
the $q$-difference operators $\mathcal{H}_l^z$ $(l=0,1,2\ldots)$ satisfy
\begin{align}
\mathcal{H}^z_l\,P_{\lambda}(z)=P_{\lambda}(z)\,g_l(t^\delta q^\lambda)
\qquad(l=0,1,2,\ldots)
\end{align}
for some symmetric polynomials $g_l(\xi)\in\mathbb{C}[\xi]^{\mathfrak{S}_n}$.  
By the Wronski type formula \eqref{eq:recW}, these polynomials 
are determined by the recurrence relation
\begin{align}
\dsum{r+s=0}{}(-1)^r (1-t^rq^s) e_r(\xi)\,g_s(\xi)=0\quad(l=1,2,\ldots).  
\end{align}
In view of 
\begin{align}
E(\xi;u)=\dsum{r=0}{n}(-u)^r e_r(\xi)=\dprod{i=1}{n}(1-u\xi_i), 
\end{align}
let us introduce the generation function
$G(\xi;u)=\dsum{l=0}{\infty}\,u^l\,g_l(\xi)$. 
Then the recurrence formula above is equivalent to the functional equation
\begin{align}
E(\xi;u)\,G(\xi;u)=E(\xi;tu)\,G(\xi; qu), 
\end{align}
namely
\begin{align}
G(\xi;u)=\dfrac{E(\xi;tu)}{E(\xi;u)}G(\xi;qu)=
\left(\dprod{i=1}{n}\dfrac{1-tu\xi_i}{1-u\xi_i}\right)G(\xi;qu).  
\end{align}
Hence we have
\begin{align}
G(\xi;u)=\dsum{l=0}{\infty} u^l g_l(\xi) =\dprod{i=1}{n}\dfrac{(tu\xi;q)_\infty}
{(u\xi_i;q)_\infty},
\end{align}
where $(u;q)_\infty=\prod_{i=0}^{\infty}(1-q^i u)$.  This means that 
\begin{align}
g_l(\xi)
=\dsum{\nu_1+\cdots+\nu_n=l}{}
\dfrac{(t;q)_{\nu_1}\cdots(t;q)_{\nu_n}}{(q;q)_{\nu_1}
\cdots(q;q)_{\nu_n}}\,\xi_1^{\nu_1}\cdots \xi_n^{\nu_n}
=
\dfrac{(t;q)_l}{(q;q)_l}
P_{(l)}(\xi)
\quad(l=0,1,2,\ldots).  
\end{align}
We introduce the generation function
\begin{align}
\mathcal{H}^z(u)=\dsum{l=0}{\infty} u^{l}\,\mathcal{H}_l^z 
=
\dsum{\mu\in\mathbb{N}^n}{}
u^{|\mu|}
\dprod{1\le i<j\le n}{}
\dfrac{q^{\mu_i}z_i-q^{\mu_j}z_j}{z_i-z_j}
\dprod{1\le i,j\le n}{}
\dfrac{(tz_i/z_j;q)_{\mu_i}}{(qz_i/z_j;q)_{\mu_i}}\,
T_{q,x}^{\mu}
\end{align}
for our $q$-difference operators $\mathcal{H}_l^z$ ($l=0,1,2,\ldots$).  
Then the argument above implies that 
\begin{align}
\mathcal{H}^z(u)\,P_{\lambda}(x)=P_{\lambda}(x)\,
\dprod{i=1}{n}
\dfrac{(ut^{n-i+1}q^{\lambda_i};q)_\infty}{(ut^{n-i}q^{\lambda_i};q)_\infty}
\end{align}
for any partition $\lambda=(\lambda_1,\ldots,\lambda_n)$ with $l(\lambda)\le n$. 
Note also that the recurrence formula of Wronski type is equivalent to 
\begin{align}
\mathcal{D}^z(u)\mathcal{H}^z(u)=\mathcal{D}^z(tu)\mathcal{H}^z(qu).  
\end{align}

Finally we give comments on 
the kernel identities for the trigonometric case.  
Consider two sets of variables $z=(z_1,\ldots,z_m)$ and $w=(w_1,\ldots.w_n)$, 
assuming that $m\ge n$.  
The Cauchy type kernel for this case is given by 
\begin{align}
\Pi(z;w)=\dprod{i=1}{m}\dprod{k=1}{n}\dfrac{(tz_iw_k;q)_\infty}{(z_iw_k;q)_\infty},
\end{align}
Then we have the kernel identities
\begin{align}\label{eq:DDPhi}
(DD)\qquad&\quad
\mathcal{D}^{z}(u) \Pi(z;w)=(u;t)_{m-n} \mathcal{D}^w(t^{m-n}u) \Pi(z;w), 
%\end{align}
%and
%\begin{align}
\\
\label{eq:HHPhi}
(HH)\qquad&\quad
\mathcal{H}^{z}(u) \Pi(z;w)=\frac{(t^{m-n}u;q)_\infty}{(u;q)_{\infty}} \mathcal{H}^w(t^{m-n}u) \Pi(z;w).  
\end{align}
By the kernel function of dual Cauchy type
\begin{align}
\qquad
\Psi(z;w)=\dprod{i=1}{m}\dprod{k=1}{n}(z_i-w_k), 
\end{align}
the two families of $q$-difference operators are exchanged as follows: 
\begin{align}\label{eq:HDPsi}
(HD)\qquad\quad
(u;q)_\infty \mathcal{H}^z(u) \Psi(z;w)=
(t^mq^nu;q)_\infty \widehat{\mathcal{D}}^w(u)\Psi(z;w)
\end{align}
where $\widehat{\mathcal{D}}^w(u)=\mathcal{D}^{(w|t,q)}(u)$ 
denotes the $q$-difference operator 
obtained from $\mathcal{D}^w(u)=\mathcal{D}^{(w|q,t)}(u)$ by exchanging $q$ and $t$.  

The three kernel identities $(DD)$, 
$(HH)$ and $(HD)$ mentioned above 
are equivalent to 
certain special cases of 
Kajihara's Euler transformation formula \cite{K2004}:  
For two sets of variables $z=(z_1,\ldots,z_m)$, $w=(w_1,\ldots,w_n)$ 
and parameters $a=(a_1,\ldots,a_m)$, $b=(b_1,\ldots,b_n)$, 
\begin{align}\label{eq:Kajihara1}
&
\frac{(u/\alpha;q)_\infty}{(u;q)_\infty}
\sum_{\mu\in\mathbb{N}^m} 
(u/\alpha)^{|\mu|}
\prod_{1\le i<j\le m}\frac{q^{\mu_i}z_i-q^{\mu_j}z_j}{z_i-z_j}
%\frac{\Delta(q^{\mu}x)}{\Delta(x)}
\prod_{i,j=1}^{m}
\frac{(a_jz_i/z_j;q)_{\mu_i}}{(qz_i/z_j;q)_{\mu_i}}
\prod_{i=1}^{m}\prod_{l=1}^{n}
\frac{(z_iw_l/b_l;q)_{\mu_i}}{(z_iw_l;q)_{\mu_i}}
\nonumber
\\
&=
\frac{(u/\beta;q)_\infty}{(u;q)_\infty}
\sum_{\nu\in\mathbb{N}^n} 
(u/\beta)^{|\nu|}
\prod_{1\le k<l\le n}\frac{q^{\nu_k}w_k-q^{\nu_l}w_l}{w_k-w_l}
%\frac{\Delta(q^{\nu}y)}{\Delta(y)}
\prod_{k,l=1}^{n}
\frac{(b_lw_k/w_l;q)_{\nu_k}}{(qw_k/w_l;q)_{\nu_k}}
\prod_{k=1}^{n}\prod_{j=1}^{m}
\frac{(w_k z_j/a_j;q)_{\nu_k}}{(w_kz_j;q)_{\nu_k}},
\end{align}
where $\alpha=a_1\cdots a_m$, $\beta=b_1\cdots b_n$. 
In fact, one can verify directly that 
these three kernel identities are equivalent to the following 
special cases of \eqref{eq:Kajihara1}, respectively: 
\begin{equation}
\begin{array}{lllll}
(DD):\quad &a_j=q^{-1} &(j=1,\ldots,m),\ \ &b_l=q^{-1}&(l=1,\ldots,n),
\\[2pt]
(HH):\quad &a_j=t &(j=1,\ldots,m),\ \ &b_l=t &(l=1,\ldots,n),
\\[2pt]
(HD):\quad &a_j=t &(j=1,\ldots,m),\ \ &b_l=q^{-1} &(l=1,\ldots,n).  
\end{array}
\end{equation}

\section*{Acknowledgment}

M.N. is grateful to the Knut and Alice Wallenberg Foundation for funding his guest professorship at KTH. 

%%%%%%%%%%%%%%%%%%%%%%%%%%%%%%%%%%%%%%%%%%%%%%%%%%%

\end{document}